\documentclass[twocolumn]{revtex4}
\usepackage{graphicx,amsmath,amsfonts}
\usepackage{color}

\def\cL{\mathcal{L}}
\def\cM{\mathcal{M}}
\def\cD{\mathcal{D}}
\def\cS{\mathcal{S}}

\def\FCW{1.0\columnwidth}
\def\Id{1\!\mathrm{l}}

\newcommand{\ket}[1]{\left| #1 \right\rangle}

\newcommand{\proj}[1]{| #1\rangle\!\langle #1 |}
\newcommand{\Tr}{\mathrm{Tr}}
\newcommand{\expect}[1]{\left\langle#1\right\rangle}

\newcommand{\LR}{\Lambda}
\newcommand{\logLR}{\lambda}

\newcommand{\reals}{\mathbb{R}}
\newcommand{\rhohat}{\hat{\rho}}
\newcommand{\rhoMLE}{\rhohat_\mathrm{MLE}}
\newcommand{\rhotrue}{\rho_0}

\begin{document}
\title{Entanglement verification with finite data}
\author{Robin Blume-Kohout$^1$, Jun O.S.\ Yin$^2$, S.\ J.\ van Enk$^{2,3}$}
\affiliation{
$^1$Perimeter Institute for Theoretical Physics, Waterloo ON N2L2Y5\\
$^2$Department of Physics and Oregon Center for Optics, University
of Oregon,
 Eugene, OR 97403\\
$^3$Institute for Quantum Information, California Institute of Technology, Pasadena, CA 91125\\
}
\date{\today}
\begin{abstract}
Suppose an experimentalist wishes to verify that his apparatus produces entangled
quantum states.  A finite amount of data cannot \emph{conclusively}
demonstrate entanglement, so drawing conclusions from real-world data
requires statistical reasoning.  We propose a reliable method to 
quantify the weight of evidence for (or against) entanglement, based
on a likelihood ratio test.  Our method is universal in that it can be applied to any sort of measurements. We demonstrate the method by applying it
to two simulated experiments on two qubits.  The first measures a single
entanglement witness, while the second performs a tomographically complete measurement.
\end{abstract}
\maketitle

\begin{figure}[t]
  \centerline{
    \includegraphics[width=\FCW]{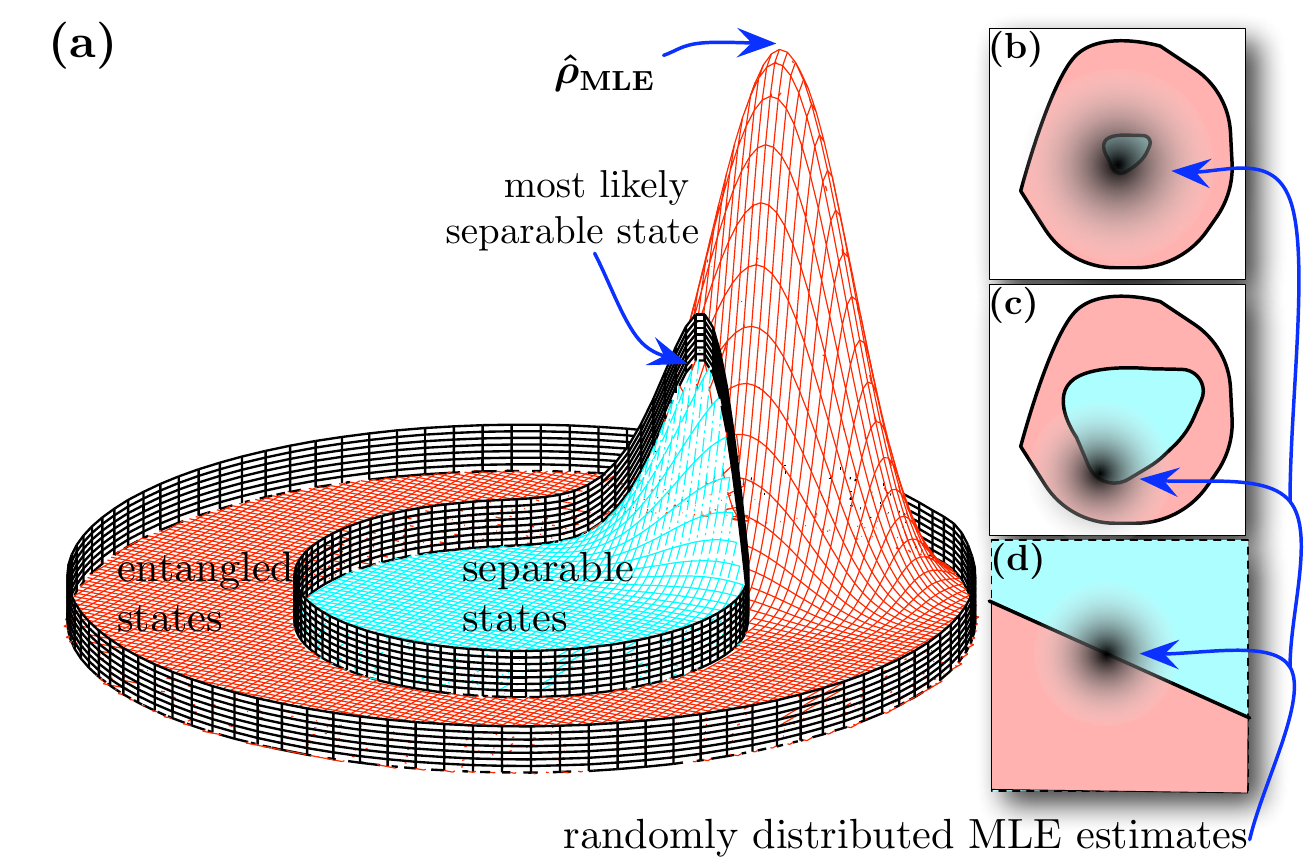}}
  \caption{\textbf{General schema of a likelihood ratio test}.  The separable states $\cS$ (cyan) are a convex subset of all states, surrounded by entangled states (red).  Data from an experiment on a state $\rho$ yield a quasiconvex likelihood function [\textbf{(a)}] with a unique maximum ($\rhoMLE$).  $\rhoMLE$ is randomly distributed around $\rho$, at a typical length scale $\delta=O(1/\sqrt{N})$.  If $\rhoMLE$ is separable then there is no evidence for entanglement, but if it's entangled (as shown), then the relative likelihoods of $\rhoMLE$ and the most likely separable state determine the weight of evidence.   Data are ``convincing'' if they are very unlikely to have been produced by a borderline separable state.  Typical likelihood ratios for such states depend on the shape of $\cS$.  In \textbf{(b)-(d)} we show three possible cases: in \textbf{(b)} $\cS$ is smaller than $\delta$ and behaves like a point; in $\textbf{(c)}$ it is of size $\delta$ and its behavior is hard to characterize; 
in $\textbf{(d)}$ it is much bigger than $\delta$ and behaves like a half-space.
}
\label{ConvexSetsFigure}
\end{figure}

Entanglement is an essential resource for quantum information processing, and producing and verifying entangled states is considered a
benchmark for quantum experiments (for a sample from the most recent experiments on a wide variety of physical systems, see \cite{exp09}).  Several methods for verifying entanglement
have been developed (for overviews, see \cite{guehne,elk}).  A bipartite state
is entangled if it is not separable, and data $\cD$ \emph{demonstrate}
entanglement if there is no separable state that could have
generated them.  As the number of data $N\to\infty$, the data are unambiguous, but
for finite $N$, only probabilistic conclusions can be drawn.  In
this Letter, we quantify exactly what can be concluded from finite
or small data sets, using a simple and efficient \emph{likelihood
ratio} test.

We demonstrate the method using two simulated experiments on two-qubit
systems \footnote{For larger systems, determining whether a given state is entangled is an NP-hard problem.  In multi-partite systems, different classes of entanglement exist, but their classification is still an open problem. Our likelihood ratio method applies to any case where the decision is {\em binary}: do the data demonstrate entanglement in a particular class or not? So in this paper, the word ``separable'' can be generalized to ``not in the desired entanglement class.''}.  The first measures just one observable, an
\emph{entanglement witness} \cite{Witness}.  The other performs a
tomographically complete measurement.  In both cases, we use likelihood
ratios to draw direct conclusions about entanglement, rather than estimating
the quantum state as an intermediate step.  A related technique for testing
violation of local realism, and based on empirical
relative entropy instead of the likelihood ratio, was proposed by van Dam et al \cite{vanDamIEEE05}
and applied by Zhang et al \cite{Zhang10}.

\vspace{0.1in}\noindent\textbf{Likelihood Ratios:} Data $\cD$ could have been
generated by any one of many i.i.d.\ states $\rho^{\otimes N}$.
Each state $\rho$ represents a theory about the system, and the
relative plausibility of different states is measured by their
\emph{likelihood} $\cL(\rho)$.  A state's likelihood is simply the
probability of the observed data \emph{given} that state,
\begin{equation}\cL(\rho) \equiv \Pr(\cD|\rho),\end{equation}
and states with higher likelihood are more plausible.  If the
\emph{most} likely state is separable, the data clearly do not
support entanglement.  If it is entangled, then we need to ask how
convincing the data are -- specifically, whether some separable
state is almost as plausible.
To judge whether there is (even just one) separable state that
fits the data, we compare the likelihoods of (i) the most likely
separable state, and (ii) the most likely of all states.
Letting $\cS$ be the set of separable states, we define
\begin{equation}
\LR \equiv \frac{\max_{\rho\in\cS}{\cL(\rho)}}{\max_{\mathrm{all\ }\rho}{\cL(\rho)}}.
\end{equation}
$\LR$ is a \emph{likelihood ratio}, and
\begin{equation}\label{L}
\logLR = -2\log \LR 
%= -2\log\frac{\max_{\rho\in\cS}\Pr(\cD|\rho)}{\max_{\mathrm{all\ }\rho}\Pr(\cD|\rho)},
\end{equation}
represents the weight of evidence in favor of entanglement
\footnote{The factor of $-2$ may seem arbitrary.  Statisticians use this
convention because $\logLR$ (as defined) is in many circumstances
a $\chi^2$ random variable (see text, below).}.
To demonstrate entanglement convincingly, an experiment must yield
a sufficiently large value for $\logLR$.  

A likelihood ratio does not assign a probability to
``$\rho$ is entangled''.  Instead, it yields a \emph{confidence level}.
We can determine what values of $\logLR$ typically result from measurements
on $\rho^{\otimes N}$, and how their distribution depends on whether $\rho$
is entangled or separable.
%An experiment that yields $\logLR$ much higher than the typical value for any
%separable state provides convincing evidence for entanglement.  More precisely,
If we measure $\logLR = \logLR_{\mathrm{exp}}$, and no separable state
produces $\logLR\geq\logLR_{\mathrm{exp}}$ with probability higher than $\epsilon$,
then we have demonstrated entanglement at the $1-\epsilon$ confidence level.
%Operationally, this means that 
If an experimentalist plans (\emph{before} taking data)
to calculate $\logLR$ and report ``$\rho$ is entangled'' \emph{only} when the
data imply $1-\epsilon$ confidence, then the probability that he erroneously
reports entanglement \footnote{Statisticians call this a ``Type I error''.
Erroneously \emph{rejecting} entanglement, even though the experiment is capable
of demonstrating entanglement (which is not the same as reporting separability),
is a ``Type II error''.  In entanglement verification one tries to avoid Type I
errors and is merely mildly unenthusiastic about Type II errors.} is at most $\epsilon$.

So, $\rho$ may be (i) entangled, (ii) separable, or (iii) on the boundary.  Boundary
states are still separable, and they are the hardest separable states to rule out.  To
demonstrate entanglement at the $1-\epsilon$ confidence level, we must show that there
is \emph{no} boundary state for which $\Pr(\logLR \geq \logLR_{\mathrm{exp}})
\geq\epsilon$.  It is difficult to make rigorous probabilistic statements about $\logLR$
for small $N$.  But as $N\to\infty$, the following analysis becomes exact, and is
generally thought to be reliable for $N\gtrsim30$ \cite{GewekeJASA80}.

\vspace{0.1in}\noindent\textbf{The distribution of $\logLR$:}  The set of
quantum states $\rho$ is a convex subset of the vector
space of trace-1 $d\times d$ Hermitian operators, $\reals^{d^2-1}$.
An entanglement-verification measurement is represented by a POVM
(positive operator-valued measure) $\cM = \{E_i\ldots E_m\}$, in which
each operator $E_k$ represents an event that occurs with probability
$p_k = \Tr E_k\rho$ (Born's rule), and each $\rho$ defines a probability
distribution $\vec{p}=\{p_1\ldots p_m\}$.  Data in which $E_k$ appeared
$n_k$ times define empirical frequencies $\vec{f} = \{f_1\ldots f_m\}$, where 
$f_k \equiv \frac{n_k}{N}$.  Both $\vec{p}$ and $\vec{f}$ can be represented
as elements of an $m$-simplex embedded in a vector space $\reals^{m-1}$.
The probabilities in $\vec{p}$ may be linearly dependent
(e.g., if $E_j+E_k=\Id$, then $p_j+p_k=1$ for all $\rho$),
and at most $d^2-1$ of them can be independent (because $\rho$ contains
only $d^2-1$ parameters).  We define $\mathrm{dim}(\cM)$ as the number of independent
probabilities.

So Born's rule defines a linear mapping from the operator space containing
quantum states into the probability space for measurement $\cM$.  
If $\mathrm{dim}(\cM)<d^2-1$, then the mapping from states
to $\vec{p}$-vectors is many-to-one, and the experiment
is completely insensitive to some parameters of $\rho$.
Ignoring these irrelevant parameters makes $\rho$ an (effectively)
$\mathrm{dim}(\cM)$-dimensional parameter.  Separable states
form a convex subset of all states (see Fig. \ref{ConvexSetsFigure}).
These sets' images in probability space are also nested convex
sets (although if $\mathrm{dim}(\cM)<d^2-1$, then some entangled
states will be indistinguishable from separable ones in this experiment).

\begin{figure}
      \includegraphics[width=\FCW]{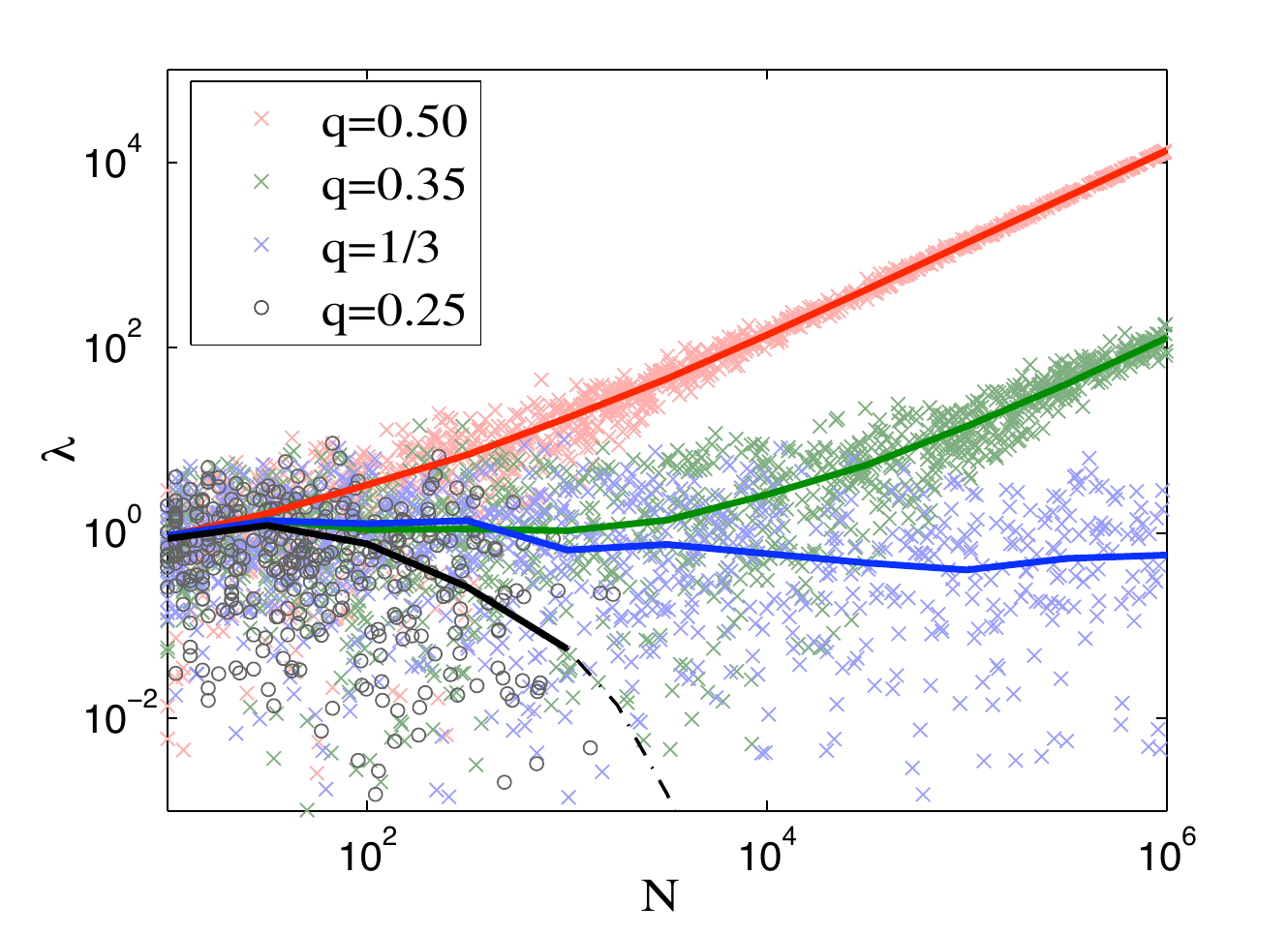}
  \caption{\textbf{Loglikelihood ratios ($\logLR$) behave dramatically differently for different states.}  1000 independent simulated tomographically complete experiments were performed, on four different Werner states -- separable, barely separable, slightly entangled, and highly entangled.  $\logLR$ is shown for each trial (points), and averaged over all 1000 trials (solid lines).  For small $N$ the experiment cannot reliably distinguish them.  As $N$ grows, it resolves shorter distances in the state space.  For entangled states, typical values of $\logLR$ increase linearly with $N$, whereas the separable state almost certainly yields $\logLR=0$ [not visible in these plots; for $\rho_{q=0.25}$ (black), all trials with more than $N\sim10^3$ measurements yielded $\logLR=0$, and the average (dashed line)plunges off the graph].  For barely separable states, $\logLR$ behaves as a semi-$\chi^2_k$ variable with $k=1$ as $N\to\infty$ (see Fig. \ref{Werner033ChiFit}).  }
\label{Werner_multiple_chifit}
\end{figure}

Suppose that $N$ copies of a state $\rhotrue$ are measured, yielding
a likelihood function $\cL(\rho)$.
$\cL(\rho)$ has a unique global maximum $\rhoMLE$.  As $N\to\infty$,
the distribution of $\rhoMLE$ approaches a Gaussian around $\rhotrue$ with
covariance tensor $\Delta$.  $\cL(\rho)$ itself is a Gaussian function
with the \emph{same} covariance matrix $\Delta$ (see note \footnote{Technically, this Gaussian ansatz is true only when $\rhotrue$ is full rank -- i.e., not on the boundary of the state set.  If $\rhotrue$ is rank-deficient, then both the distribution of $\rhoMLE$ and $\cL(\rho)$ itself are typically truncated by the boundary.  However, the analysis remains valid (as $N\to\infty$) except if $\rhotrue$ is simultaneously rank-deficient \emph{and} on the boundary between separable and entangled states.}).  This defines a characteristic
length scale $\delta = |\Delta|_2$ that scales as $\delta=O(1/\sqrt{N})$.
We can use $\Delta$ to define a stretched Euclidean metric
\begin{equation}
d(\rho_1,\rho_2) = \sqrt{\Tr\left[(\rho_1-\rho_2)\Delta^{-1}(\rho_1-\rho_2)\right]}.
\end{equation}
Using this metric, $\rhoMLE$ is univariate Gaussian distributed around $\rhotrue$, and
\begin{equation} \label{eq:logL_is_distance}
\log\cL(\rho) = -\frac{d(\rho,\rhoMLE)^2}{2}.
\end{equation}
Thus, $\logLR$ is determined entirely by $d(\rhoMLE,\cS)$, the distance from $\rhoMLE$
to the separable set $\cS$.  If $\rhotrue$ is demonstrably entangled, then $\logLR$ will grow
proportional to $N$ -- but if it is indistinguishable from a separable state, then $\logLR$ will converge almost certainly to zero (see Figure \ref{Werner_multiple_chifit}).

%and $d(\rhotrue,\rhoMLE)^2$ is a $\chi^2$ random variable with $\mathrm{dim}(\cM)$ degrees of freedom.

When $\rhotrue$ is on the boundary, $\logLR$ neither grows with $N$ nor converges to zero, but continues to fluctuate as $N\to\infty$.  Its distribution is controlled by the shape and radius of $\cS$, e.g.:
\begin{enumerate}
\item If $\cS$ is small w/r.t. $\delta$, it behaves like a point (see Figure \ref{ConvexSetsFigure}b).  Then $d(\rhoMLE,\cS)\approx d(\rhoMLE,\rhotrue)$, $\logLR = -2\log\left(\cL_{max}/\cL(\rhotrue)\right) = d(\rho,\rhoMLE)^2$, and so $\lambda$ is a $\chi^2$ random variable with $\mathrm{dim}(\cM)$ degrees of freedom (a.k.a. a $\chi^2_{\mathrm{dim}(\cM)}$ variable).
\item If $\cS$ is much larger than $\delta$, then it behaves like a half-space (see Figure \ref{ConvexSetsFigure}d and note \footnote{As long as the boundary of $\cS$ is differentiable at $\rhotrue$.}).  If $\cS$ were a $k$-dimensional hyperplane, $\logLR$ would be a $\chi^2_{\mathrm{dim}(\cM)-k}$ variable.  A halfspace behaves like a hyperplane of dimension $(\mathrm{dim}(\cM)-1)$, except with probability $\frac12$, $\rhoMLE$ is separable.  Thus, $\logLR$ is what we will call a \emph{semi-$\chi^2_1$} variable:  it equals zero with probability $\frac12$, and is $\chi^2_1$-distributed otherwise.
\end{enumerate}
As $N\to\infty$, case (2) applies.  For small $N$, however, the real situation is somewhere in between (see Figure \ref{ConvexSetsFigure}c).  $\cS$ may be small, and its boundary may be sharply curved, increasing $\logLR$.  In the absence of a detailed understanding of $\cS$'s shape, case (1) provides the best rigorous upper bound on $\logLR$.  Its cumulative distribution is upper bounded by that of a $\chi^2_{\mathrm{dim}(\cM)}$ variable -- i.e., $\Pr(\logLR > x)$ is no greater than it would be if $\logLR$ was a $\chi^2_{\mathrm{dim}(\cM)}$ variable.  As $N\to\infty$, the more optimistic semi-$\chi^2_1$ ansatz is valid -- but only if we know that $N$ is ``large enough''.

A $\chi^2_k$ variable has expected value $k$, and higher values are exponentially suppressed.  So $\logLR \gg \mathrm{dim}(\cM)$ is sufficient to demonstrate entanglement at a high confidence level.  This implies a tradeoff between an experiment's \emph{power} (ability to identify many entangled states) and its \emph{efficiency} (ability to do so rapidly).  Powerful experiments have large dimension -- e.g., a tomographically complete measurement can identify \emph{any} entangled state, but has $\mathrm{dim}(\cM)=d^2-1$.  This comes at a price; experiments with large dimension are potentially much more prone to spurious large values of $\logLR$, so more data is required to achieve conclusive results [$\logLR\gg\mathrm{\dim}(\cM)$].  Conversely, an entanglement witness (see below) is targeted at a particular state, but it can rapidly and conclusively demonstrate entanglement.

% FIGURE 2

\vspace{0.1in}\noindent\textbf{Implementation:}
Computing $\logLR$ involves maximizing $\cL(\rho)$ over two convex
sets (the set of all states, and the set $\cS$ of separable states).
$\cL(\rho)$ is log-convex, so in principle this is a convex program.

Testing separability is NP-hard, so efficient
minimization over $\rho\in\cS$ is  impossible in general.  But for
two qubits, the positive partial transpose (PPT) criterion
perfectly characterizes entanglement, and $\logLR$ can be calculated easily 
(see examples below).  For larger systems, $\cS$ can be bounded by simpler convex sets, as
$\cS_- \subset \cS \subset \cS_+,$
(e.g., $\cS_+ =\ $ PPT states, and $\cS_- =\ $ convex
combinations of specific product states).  Maximizing $\cL(\rho)$ over
$\cS_+$ and $\cS_-$ yields bounds on $\max_{\rho\in\cS}{\cL(\rho)}$,
which may (depending on how wisely the bounding sets were chosen) be
tight enough to confirm or deny entanglement.

\noindent\textbf{Examples:}  To demonstrate the likelihood ratio test, we simulate two different experiments on two
qubits.  We imagine an experimentalist
trying to produce the singlet state
$\ket{\Psi}$, and producing instead
a \emph{Werner state} \cite{Werner},
\begin{equation}
\rho_q=q\Pi_{{\rm singlet}}+(1-q)\openone/4,
\end{equation}
where $\Pi_{\rm singlet} = \proj{\Psi}$.  Werner states are separable
when $q\leq 1/3$, and entangled otherwise.  The experimentalist's
repeated preparations are assumed to be independently and identically
distributed (i.i.d.) \cite{renner}.

% FIGURE 3
% Insert the figure here so that the layout of the last page would work out 
\begin{figure}[t]
  \centerline{
    \includegraphics[width=\FCW]{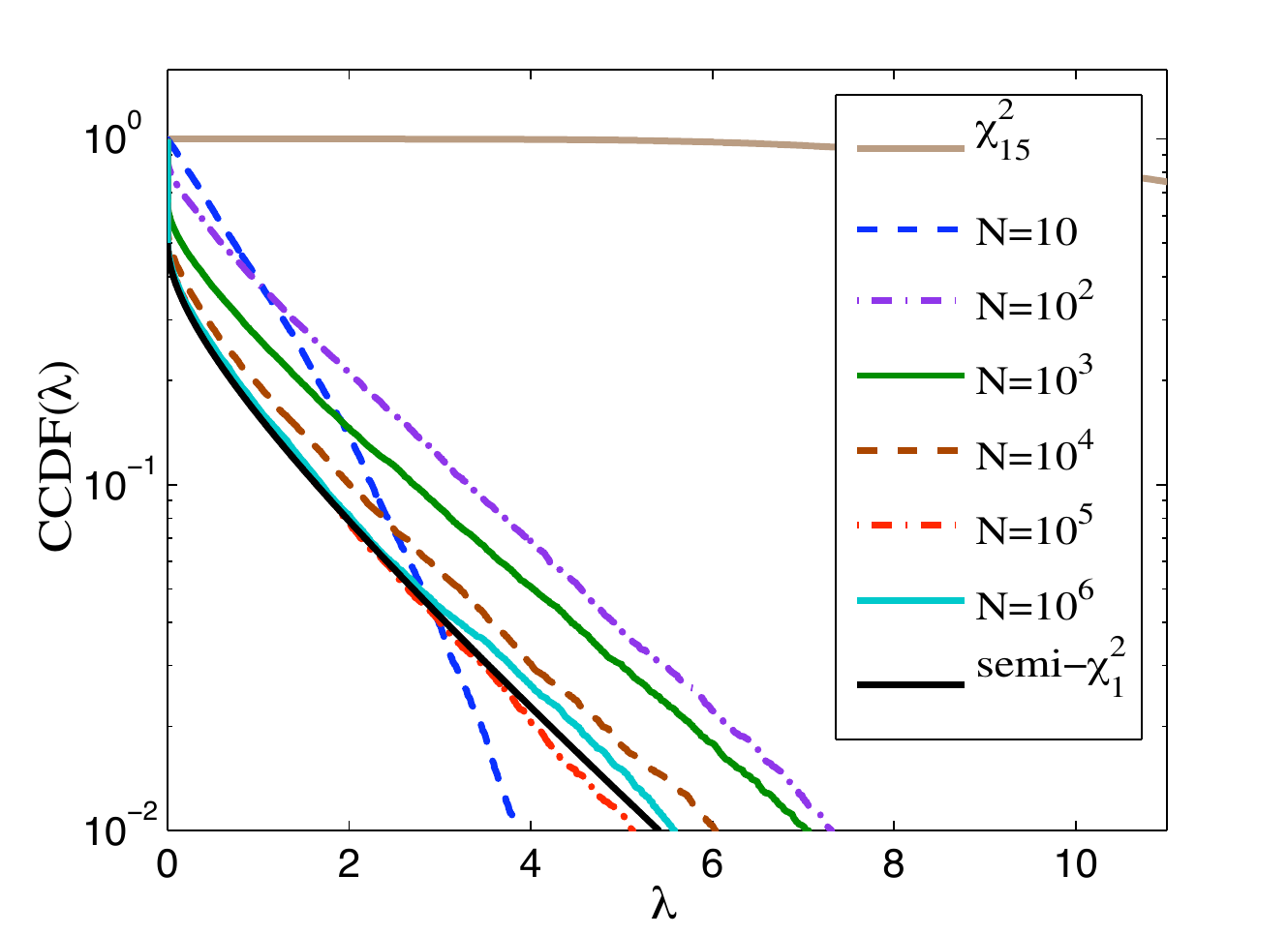}}
   \caption{\textbf{Distribution of $\logLR$ for a SIC-POVM experiment}.  We show the empirical \emph{complementary cumulative distribution function} of $\logLR$, $CCDF(\logLR_c) = \Pr(\logLR>\logLR_c)$, for the state $\rho_{q=1/3}$ and simulated datasets of size $N=\{10\ldots10^6\}$.  The CCDF is used to compute confidence levels -- e.g., to report entanglement at the 95\% confidence level, it is necessary to observe $\logLR$ such that $CCDF(\logLR) < 0.05$.  For this particular state, the chance of a zero $\lambda$ approaches $50\%$ as $N$ increases. For each $N$, $CCDF(\logLR_c)$ was based on roughly $10^4$ data points from independent trials, each of which generated a value of $\logLR$ from $N$ tomographically complete measurements on $\rho_{q=1/3}$.  We also show CCDFs for a semi-$\chi^2_1$ variable and a $\chi^2_{\dim(\cM)}=\chi^2_{15}$ variable. The semi-$\chi^2_1$ ansatz is good for large $N$, but unreliable for small $N$ (yielding too many false positives), while the $\chi^2_{15}$ ansatz is very conservative.}
\label{Werner033ChiFit}
\end{figure}

\vspace{0.1in}\noindent\textbf{Witness data:} The simplest way to test for entanglement is to repeatedly measure a
single \emph{entanglement witness} \cite{Witness, guehne}. An optimal witness for Werner states is
$W=\openone/2-\Pi_{{\rm singlet}}$.
Measuring $W$ yields one of two outcomes -- 
``yes'' or ``no'' -- corresponding to POVM (positive-operator valued
measure) elements $\{\Pi_{\rm singlet},\openone-\Pi_{\rm singlet}\}$.  The
probability of a ``yes'' outcome is given by Born's rule as
$p={\rm Tr} \rho\Pi_{{\rm singlet}},$
so $p$ completely characterizes a state $\rho$ for the purposes of
this experiment.  The data from $N$ measurements is fully characterized by the frequency
of ``yes'' results, $f=n_{\rm ``yes''}/N$.  As $N\rightarrow \infty$,
$f>\frac12$ represents definitive proof that $\expect{W}<0$, and therefore
that $\rho$ is entangled.  For finite $N$, $f\leq\frac12$ means that a separable
state fits as well as any other, so there is no case for
entanglement.  When $f>\frac12$, our likelihood ratio quantifies
the weight of the evidence for entanglement. 
The likelihood function depends only on $p$, as
\begin{eqnarray}
\cL(\rho) &=& \cL(p) = \mathrm{Pr}(\vec{f}|p) = p^{Nf}(1-p)^{N(1-f)}\nonumber\\
&=& {\rm e}^{-N\left(-f\log p - (1-f)\log(1-p)\right)},
\end{eqnarray}
making this a single-parameter problem.
The maximum likelihood, attained at $p=f$, is $\cL_{\rm max}  = {\rm e}^{-N H(f)}$,
expressed in terms of the data's \emph{empirical entropy},
\begin{equation}
H(f)=-f\log f-(1-f)\log(1-f).
\end{equation}
If $f>\frac12$, the most likely
separable state has $p=\frac12$, so
that $
\cL_{\rm sep}  = 2^{-N},
$
which yields 
\begin{equation}
\logLR = -2\log \frac{\cL_{\rm sep}}{\cL_{\rm max}} = 2N\left[\log(2) -H(f)\right].
\end{equation}
Our numerical explorations (not shown here) confirm that for a barely-separable Werner state, $\logLR$ behaves as a semi-$\chi^2_1$ variable, even for $N$ as low as $20$.

%\textbf{Since $\tilde{P}(\logLR)=P(f)\cdot(\diff \logLR/\diff f)^{-1}$, we get a distribution of $\logLR$ as a function of $f$:}\begin{eqnarray}
%  \tilde{P}(\logLR)=&&\frac{\Gamma(N+1)}{\Gamma(Nf+1)\Gamma(N(1-f)+1)}p^{Nf}(1-p)^{N(1-f)}\nonumber\\
%  &&\cdot N\log\frac{f}{1-f},
%\end{eqnarray}
%\textbf{where $\Gamma(N+1)=N!$ when $N$ is a positive integer. The distribution of $\logLR$ is shown in Fig.\ref{Werner033_witness_chifit}}.

\vspace{0.1in}\noindent\textbf{Tomographically complete data:} 
Many entanglement-verification experiments measure a tomographically
complete set of observables on a finite-dimensional system (with a heroic example being tomography on 8 ions in an ion trap \cite{blatt}).  Such data
identify $\rho$ uniquely as $N\to\infty$, so one can
determine with certainty whether $\rho$ is entangled (modulo the computational
difficulties in determining whether a specified $\rho$ is separable).  
Analyzing finite data is more complicated than in the witness example, for
the data constrain a multidimensional parameter space.  Ad-hoc techniques are
unreliable, and the likelihood ratio test comes into its own.

We consider an apparatus that applies a SIC (symmetric informationally
complete)-POVM \cite{sicpovm} to each of our two qubits, independently.
This measurement (not to be confused with a 4-dimensional SIC-POVM) is
tomographically complete, has $4\times4=16$ outcomes, and yields 15
independent frequencies.  Unlike $W$, it has no special relationship to Werner
states, so \emph{any} entangled $\rho$ will yield overwhelmingly
convincing data as $N\to\infty$.

We repeatedly simulated 
$N=10\ldots 10^6$ measurements on a barely-separable Werner state ($\rho_{q=1/3}$),
and compared the empirical distribution of $\logLR$ to those of semi-$\chi^2_1$ and $\chi^2_{15}$
random variables (see Figure \ref{Werner033ChiFit}).  As $N$ gets large, 
$\logLR$ becomes indistinguishable from a semi-$\chi^2_1$ variable.
For smaller $N$, this ansatz is too optimistic (and would produce excessive false positives),
but the $\chi^2_{d^2-1}$ ansatz is wildly overcautious.  We found that for small $N$,
$\logLR$ behaves like a semi-$\chi^2_D$ variable, with $D$ a bit larger than 1 (e.g. 
$D\approx1.6$ for $N=100$).

\vspace{0.1in}\noindent\textbf{Conclusions:}
Entanglement verification is easy when $N\to\infty$.
%The data either confirm entanglement, confirm separability, or (for 
%tomographically incomplete measurements) are definitely
%inconclusive.
In practice, $N$ is finite and data are
never conclusive.  Likelihood ratios provide a simple, reliable test
of significance that can be applied to \emph{any} experimental data.
Large values of $\logLR$ are \emph{very} unlikely to be generated
by any separable state, but the hardest separable states to rule out
are on the boundary.  For such states, theory predicts (and our
numerics confirm) that $\logLR$ behaves like a semi-$\chi^2$ random variable.
If the underlying state is separable, $\Pr(\logLR>x)$ can be upper bounded
using a $\chi^2_{\mathrm{dim}(\cM)}$ distribution, scaling as $e^{-x}$ for large $x$.
For entangled states, $\logLR$ grows linearly with $N$, and will thus
rapidly become distinguishable from any separable state.

%In closing, we note three compelling questions for future work.  First, can
%$\cS$, the set of separable states, be tightly approximated by simpler
%bounding sets $\cS_-$ and $\cS_+$?  Good bounds on $\lambda$ for larger systems
%will require such bounding sets.  Second, how can we characterize the
%distribution of $\logLR$ for small $N$ and barely separable $\rho$?  And finally,
%can alternative methods (especially Bayes factors) efficiently provide better results
%when correctly tuned?


\begin{thebibliography}{99}
\bibitem{exp09}L. Hofstetter {\em et al.}, Nature {\bf 461}, 960 (2009);
M. Ansmann {\em et al.}, Nature {\bf 461}, 504 (2009);
E. Amselem and M. Bourennane, Nature Physics {\bf 5}, 748 (2009);
P. Bohi {\em et al.}, Nature Physics {\bf 5}, 592 (2009);
L. DiCarlo {\em et al.}, Nature {\bf 460}, 240 (2009);
J. Janousek {\em et al.}, Nature Photonics {\bf 3}, 399 (2009);
J.D. Jost {\em et al.}, Nature {\bf 459}, 683 (2009);
J.C.F. Matthews {\em et al.}, Nature Photonics {\bf 3}, 346 (2009);
A. Fedrizzi {\em et al.}, Nature Physics {\bf 5}, 389 (2009);
A. Ourjoumtsev {\em et al.}, Nature Physics {\bf 5}, 189 (2009);
A. S. Coelho {\em et al.}, 
Science {\bf 326},  823 (2009);
Scott B. Papp {\em et al.}, 
Science {\bf 324} 764 (2009);
Ryo Okamoto {\em et al.}, 
Science {\bf 323},  483 (2009);
S. Olmschenk {\em et al.},
Science {\bf 323}, 486 (2009).



\bibitem{guehne}O. G\"uhne and G. Toth,
Physics Reports {\bf 474}, 1 (2009).


\bibitem{elk}
S.J. van Enk, N. Lutkenhaus, and H.J. Kimble, Phys.
Rev. A {\bf 75}, 052318 (2007)

\bibitem{Witness}
M.\ Horodecki, P. Horodecki, and R. Horodecki, Phys. Lett. A {\bf 223}, 1 (1996);
B. Terhal, Physics Letters A {\bf  271},  319 (2000);
P.\ Hyllus {\it et al.}, Phys. Rev. A {\bf 72}, 012321 (2005).

\bibitem{vanDamIEEE05}
W.\ van Dam, R.\ D.\ Gill, and P.\ D.\ Grunwald, IEEE Trans. Inf. Th. {\bf 51}, 2812 (2005).

\bibitem{Zhang10}
Y.\ Zhang, E.\ Knill, and S.\ Glancy, \texttt{arxiv:1001.1750} (2010).

\bibitem{GewekeJASA80}J.~F. Geweke and K.~J. Singleton, J. Am. Stat. Assoc., {\bf 75}, 133 (1980).

\bibitem{Werner}
R.F.\ Werner, Phys. Rev. A {\bf 40}, 4277 (1989).

\bibitem{renner}R. Renner, Nature Physics {\bf 3}, 645 (2007).

\bibitem{sicpovm}
J. M. Renes, R. Blume-Kohout, A. J. Scott, and C. M. Caves, Journal of Mathematical Physics, {\bf 45}, 2171 (2004).

\bibitem{blatt}H.~H\"affner {\em et al.} Nature {\bf 438}, 643 (2005).

%\bibitem{chi}For example, see W. Feller, {\em An Introduction to Probability Theory and Its Applications} New York: Wiley (1971); P.G. Hoel,  {\em Introduction to Mathematical Statistics} New York: Wiley, (1962).
    
% \bibitem{ChernoffAnnMathStat56}H. Chernoff, Ann. Math. Stat., {\bf 25}, 573 (1956).

%\bibitem{Lehmann06}E.~L. Lehmann, IMS Lecture Notes-Monograph Series, {\bf 49}, 1 (2006).

%\bibitem{toby}See Toby Cubitt's code for generating random density matrices at:{\tt http://www.dr-qubit.org/matlab.php\#quantinf}

%\bibitem{B}For example, see H. Jeffreys, {\em The Theory of Probability}, Oxford (1961).

%\bibitem{MH}  B.A. Berg, {\em Markov Chain Monte Carlo Simulations and Their Statistical Analysis}, Singapore, World Scientific (2004);
 %   C.P. Robert  and G. Casella, {\em Monte Carlo Statistical Methods}, New York, Springer (2004).

\end{thebibliography}
\end{document}